\documentclass[a4paper,fleqn,usenatbib]{mnras}

\usepackage{graphicx}	
\usepackage{amsmath}	
\usepackage{amssymb}	
\usepackage{multicol}        
\usepackage{bm}		
\usepackage{pdflscape}	
\usepackage[T1]{fontenc}
\usepackage{ae,aecompl}
\usepackage{graphicx}	
\usepackage{amsmath}	
\usepackage{amssymb}	
\usepackage{lscape}
 
\usepackage[T1]{fontenc}
\usepackage{ae,aecompl}

\usepackage{newtxtext,newtxmath}

\title[]{Does the Alfv\'{e}n wave disrupt the large-scale magnetic cloud structure?}

\author[Anil Raghav \& Ankita Kule]{
Anil N. Raghav,$^{1}$\thanks{E-mail: raghavanil1984@gmail.com}
 and Ankita Kule,$^{1}$
\\
$^{1}$University Department of Physics, University of Mumbai, Vidyanagari, Santacruz (E), Mumbai-400098, India\\
}

\date{Accepted XXX. Received YYY; in original form ZZZ}

\pubyear{2018}

\begin{document}
\label{firstpage}
\pagerange{\pageref{firstpage}--\pageref{lastpage}}
\maketitle

\begin{abstract}
Alfv\'{e}n waves are primal and pervasive in space plasmas and significantly contributes to microscale fluctuations in the solar wind and some heliospheric processes. 
 Here, we demonstrate the first observable distinct feature of Alfv\'{e}n wave while propagating from magnetic cloud to trailing solar wind. The Wal\'{e}n test 
  is used to confirm their presence in selected regions. The amplitude ratio of inward to outward Alfv\'{e}n waves 
 is employed to establish their flow direction. The dominant inward flow is observed in magnetic cloud whereas trailing solar wind shows the dominant outward flow of Alfv\'{e}n waves. The observed reduction in Wal\'{e}n slope and correlation coefficient within magnetic cloud suggest (i) the simultaneous presence of an inward \& outward Alfv\'{e}n waves and/or (ii) a possibility of magnetic reconnection and/or (iii) development of thermal anisotropy and/or (iv) dissipation of Alfv\'{e}nic fluctuations. The study implies that either the Alfv\'{e}n waves dissipate in the magnetic cloud or its presence can lead to disruption of the magnetic cloud structure. 
\end{abstract}




\begingroup
\let\clearpage\relax
\endgroup
\newpage

\section{Introduction}

Alfv\'{e}n waves \citep{alfven1942existence} \textit{i.e.} incompressible magnetohydrodynamics (MHD) waves are the primitive oscillation in a magnetized plasma which has paramount importance
both in astrophysical and laboratory plasmas  \citep{foote1979hydromagnetic,cramer2011physics}.  These waves play a crucial role across various regimes of plasma physics, for example, (i) magnetized
turbulence phenomenologies \citep{goldreich1995toward,ng1996interaction,PhysRevLett.96.115002}, (ii)the solar wind and its interaction with Earth \citep{eastwood2005foreshock,ofman2010wave,bruno2013solar,yang2013alfven,burlaga1971hydromagnetic}, (iii) interaction of magnetic large-scale structures  \citep{raghav2018first},  (iv) the solar corona \citep{marsch2006kinetic,mcintosh2011alfvenic}, (v) solar and stellar interiors \citep{gizon2008helioseismology}, (vi) cosmic-ray transport \citep{schlickeiser2015cosmic}, (vii) astrophysical disks \citep{quataert1999turbulence}, and (viii) magnetic fusion \citep{heidbrink2008basic} etc. This fascinating multitudinous applications result in an intense study of Alfv\'{e}n waves which  intriguingly attracting a lot of interest in space and solar physics \citep{cramer2011physics}.

In Alfv\'{e}n waves, fluid velocity and magnetic field are oscillating together such that it can generate a wave propagating along the direction of the magnetic tension force. This leads to the obvious characteristic for identifying Alfv\'{e}n waves i.e. the well-correlated changes in magnetic field $B$ and plasma velocity $V$, described by the Wal\'{e}n relation \citep{walen1944theory,hudson1971rotational} as,
\begin{equation}
  V_A = \pm A~ \frac{B}{\sqrt{\mu_0 \rho}}~~~~~~~
  \text{and}~~~~~
  A = \sqrt{1 - \frac{\mu_0(P_\parallel - P_\perp)}{B^2}}
\end{equation} 
where $B$ is magnetic field vector and $\rho$ is proton mass density.
The $\pm$ denotes the sign of wave vector i.e propagation directions, parallel (-) and anti-parallel (+) to the background magnetic field $B_0$. $A$ is the thermal anisotropy parameter, 
 $P_\parallel$ and $P_\perp$ are the thermal pressures parallel and perpendicular to the $(B_0)$, respectively. In the solar wind near $1~ AU$, the thermal anisotropy is considered to be non-significant, and $A$ is often assumed to be $1~$\citep{burlaga1971nature,yang2016observational}. 

The fluctuations $\Delta B$ in $B$ can be obtained by subtracting average value of $B$ from each measured values. Therefore, the fluctuations in Alfv\'{e}n velocity is estimated as
\begin{equation}
\Delta V_A = \frac{\Delta B}{\sqrt{\mu_0 \rho}} ~~~~~~~
  \text{and}~~~~~
  \Delta V = |R_{W}| \Delta V_{A}
\end{equation}
Furthermore, the fluctuations of proton flow velocity $\Delta V$ are determined by subtracting averaged proton flow velocity from measured values. The significant correlation between each respective component of $\Delta V_A$ and $\Delta V$ indicate the presence of Alfv\'{e}n wave. The linear relation between $\Delta V_A$ and $\Delta V$ express the Wal\'{e}n slope ($R_w$). 

Yang et al. (2016) proposed an analytical relation between Wal\'{e}n slope and theoretical estimate of the amplitude ratio ($R_{v_A}$) of inward to outward waves assuming that the observed Alfv\'{e}n waves are their superposition. A measure of the amplitude ratio of inward to outward Alfv\'{e}n waves is express as, 
\begin{equation}
\text{for}~~~~~B_{x} < 0,~~~~~~~~~~~~~~~~~~~
R_{{V}_{A}} = \frac{1 + R_{W}}{1 - R_{W}}
\end{equation}
 and
\begin{equation}
\text{for}~~~~~B_{x} > 0,~~~~~~~~~~~~~~~~~~~
R_{{V}_{A}} = \frac{1 - R_{W}}{1 + R_{W}}
\end{equation}
in GSE-coordinate system. The $R_{{V}_{A}}<1$ for outward Alfv\'{e}n waves, whereas $R_{{V}_{A}}> 1$ for inward Alfv\'{e}n waves \citep{yang2016observational}. It is expected that the Wal\'{e}n slope should be 1, however, most of the past observational studies show $R_W$ varies from 0.28 to 0.99 \citep{belcher1971large,goldstein1995alfven,gosling2011pulsed,chao2014walen}. The smaller (sub-unity) value of Wal\'{e}n slope in the solar wind plasma is still unresolved problem in space physics \citep{neugebauer2006comment}. Moreover, how Alfv\'{e}n wave properties transformed while propagating from one particular region to other is hardly studied. In fact, Alfv\'{e}n waves are rarely observed in the magnetic cloud even though their expectations are high \citep{gosling2010torsional,raghav2018first}. It is important to study the influence of Alfv\'{e}n waves on the dynamic evolution of magnetic cloud of coronal mass ejection (CME). Besides this, the energy flow of Alfv\'{e}n waves  within the magnetic cloud and to its following solar wind are intriguing problems. Here, we demonstrate the first \textit{in-situ} observational features of Alfv\'{e}n wave while propagating from magnetic cloud to trailing solar wind and put some light on the aforementioned issues and its implications.

\section{Event selection \& Observations}

\begin{figure*}
\begin{center}

\includegraphics[width=1\textwidth]{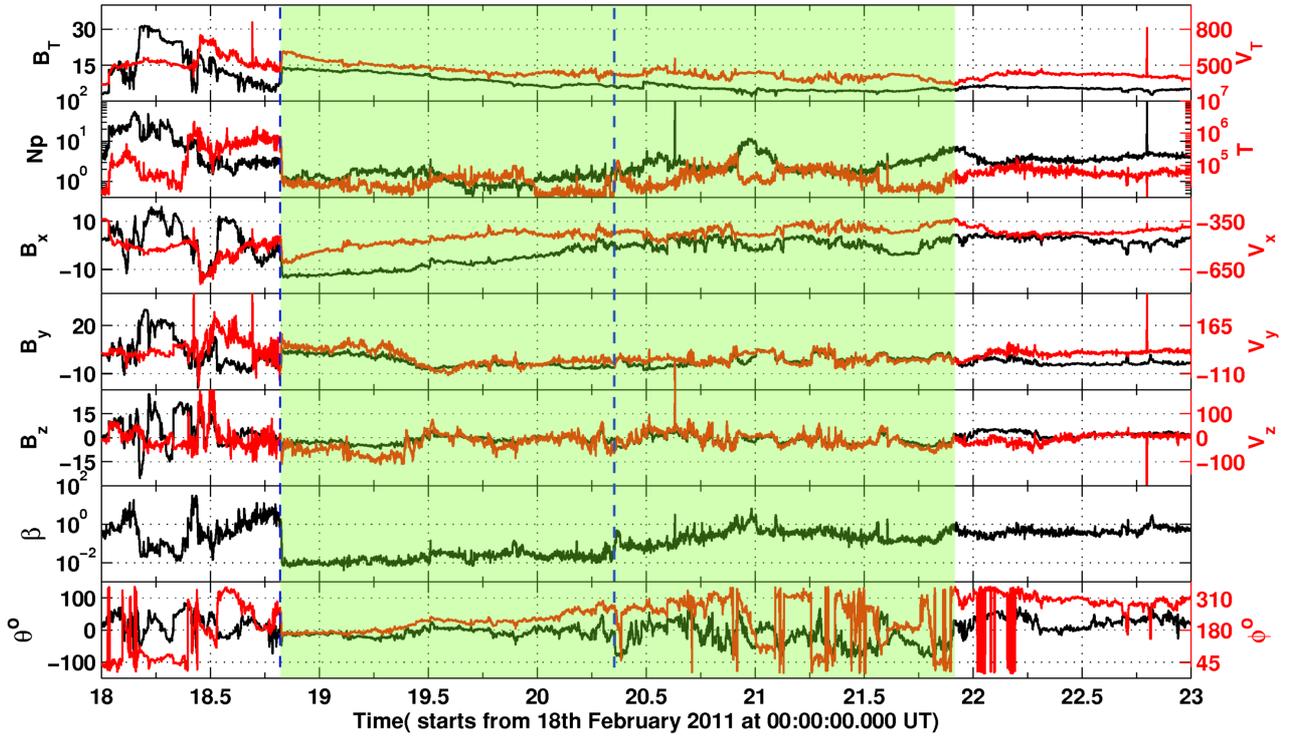}
\caption{Wind observation of complex CME-CME interaction event on 18-20 February 2011 ( time cadence of 92 sec). The top panel shows total interplanetary field strength IMF ($|B_T|$) and total solar wind $(V_T)$. The $2^{nd}$ panel shows plasma proton density and temperature. The $3^{rd}$, $4^{th}$ and $5^{th}$ panel from top show IMF components $(B_x, B_y, B_z)$ and solar wind components $(V_x, V_y, V_z)$ respectively. The $6^{th}$ panel shows plasma beta and bottom panel shows IMF orientation $(\Phi, \Theta)$. All observations are in GSE coordinate system. The blue vertical dashed lines presents the magnetic cloud boundaries. The green shade depicts the possibility of Alfv\'{e}nic fluctuations based on the visual inspection of the $3^{rd}$, $4^{th}$ and $5^{th}$ panels.
}
\label{fig:IP}
\end{center}
\end{figure*}

Here, we study CME-CME interaction event which was provoked by interacting en-route multiple-CME's erupted on $13^{th} $, $14^{th}$, and $15^{th}$ February 2011.  The same event has been investigated in past to study: 1) their interaction corresponding to different position angles \citep{temmer2014asymmetry}, 2) their geometrical properties and the coefficient of restitution for the head-on collision scenario and their geomagnetic response \citep{mishra2014morphological}, 3) corresponding Forbush decrease phenomena  \citep{marivcic2014kinematics,raghav2014quantitative,raghav2017forbush}, 4) the presence of Alfv\'{e}n waves in interacting region \citep{raghav2018first}, 5) their geo-effective response\citep{raghav2018torsional}. The \textit{in-situ} observation of the selected event in GSE-coordinate system with time cadence of 92 sec is shown in Figure ~\ref{fig:IP}. A complex magnetic structure is observed in WIND satellite data at 1 AU. The region between two blue vertical dashed lines shows low $\beta$ and plasma temperature, constant density, and the slow temporal variations in $\theta$ $\&$ $\phi$. This implies the presence of a magnetic cloud or magnetic cloud-like structure \citep{marivcic2014kinematics,raghav2018first,raghav2018torsional} .
Raghav et al.(2018) separated the green shaded region into two regions viz. Magnetic cloud and it's trailing solar wind. The Wal\'{e}n test was applied and the correlation and the linear relation between the components of $\Delta V_A$ and $\Delta V$ for both the region 1 and 2 were studied. In the magnetic cloud, the Wal\'{e}n slopes for x, y and z components were 0.57, 0.64 and 0.78 and the correlation coefficient ($R$) between Alfv\'{e}n and proton flow velocity vector components were noted as 0.84, 0.91, and 0.92 respectively. In trailing solar wind region, the Wal\'{e}n slopes for x, y and z components were 0.34, 0.36 and 0.36 and the correlation coefficients were 0.34, 0.47, and 0.55 respectively. The estimated values of the Wal\'{e}n slopes and the correlation coefficients are poor for Alfv\'{e}n waves in trailing solar wind as compared to magnetic clouds region \citep{raghav2018torsional}.

\begin{figure*}
\begin{center}
\includegraphics[width=1 \textwidth]{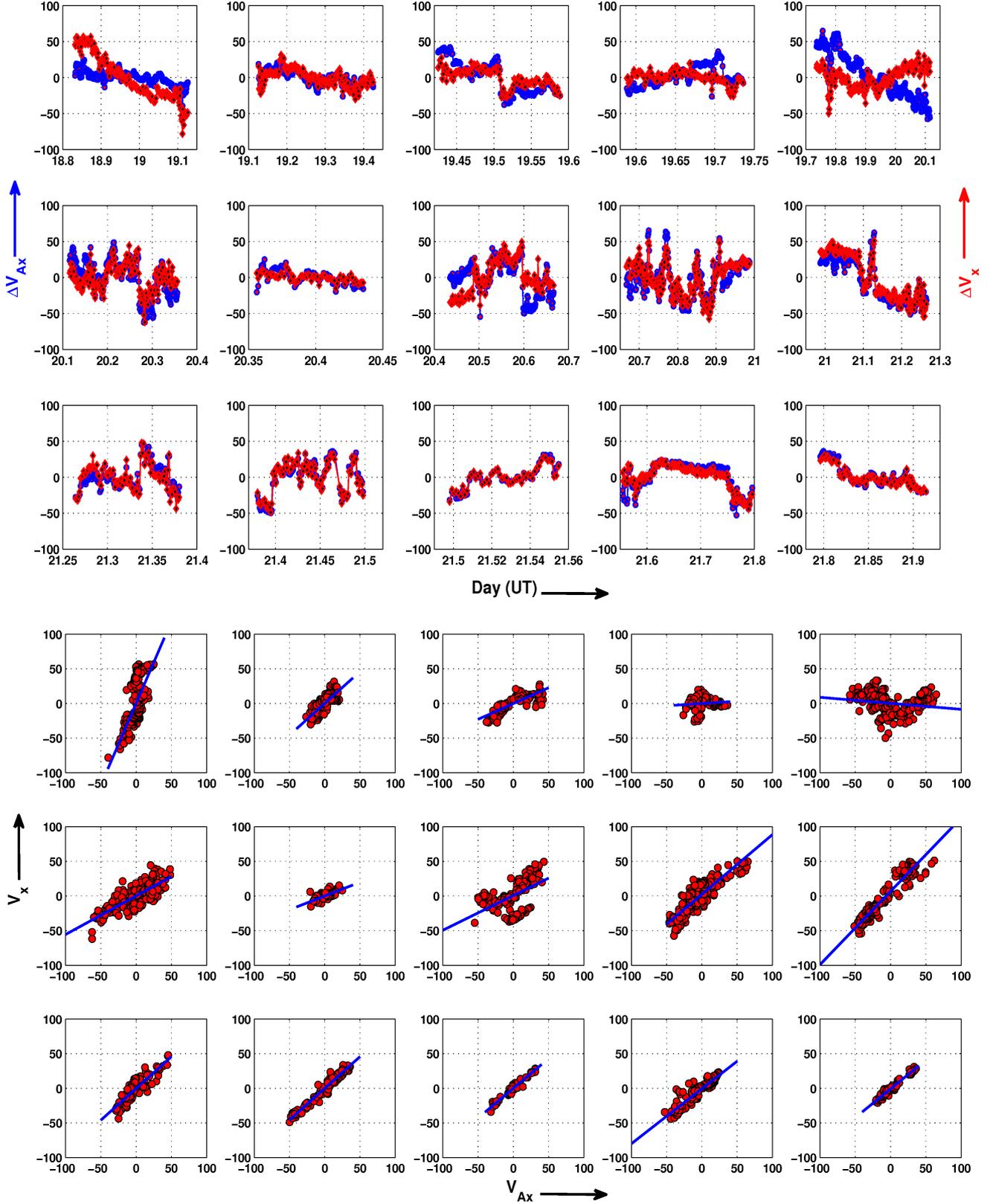}
\caption{Top right 15 panels illustrate relative fluctuation of  Alfv\'{e}n velocity vector $\Delta {V_{Ax}}$ (blue circles with line) and that of respective proton flow velocity vector $\Delta V_x$ (red circles with line). The scattered black circles with red filling are visual correlation and the linear relation between $\Delta V_{Ax}$ and $\Delta V_x$ for the aforementioned 15 subregions with same order of plotting. The estimated slopes and correlation coefficient for each subregion are shown in Table~\ref{tab:1}.  (time cadence of 92 sec).}
\end{center}
\label{fig:vx_vax}
\end{figure*}

\begin{figure*}
\includegraphics[width=1 \textwidth]{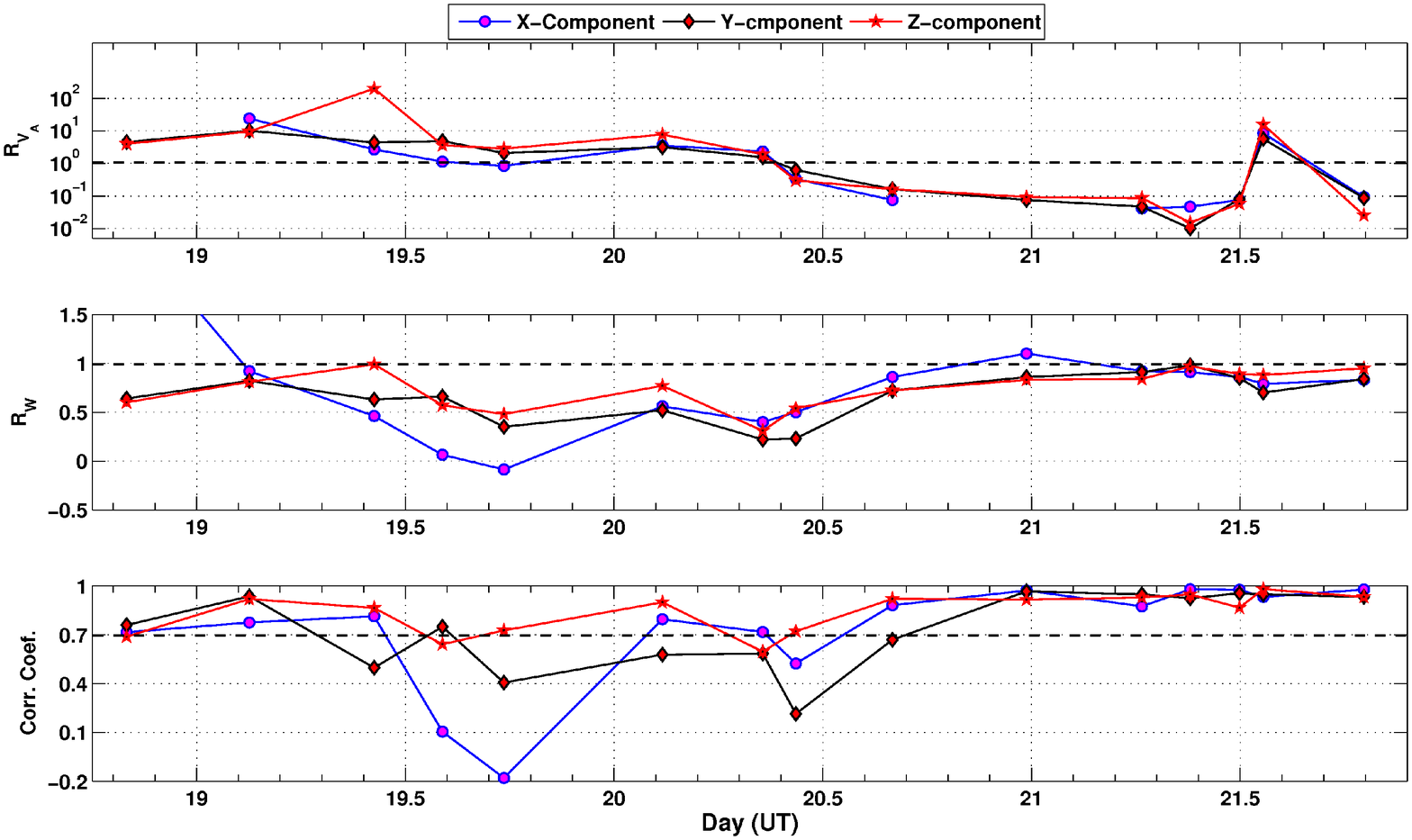}
\caption{The top panel demonstrate the variations of the amplitude ratio ($R_{v_A}$) of inward to outward Alfv\'{e}n waves, the middle panel illustrates the variation of Wal\'{e}n slope and the bottom panel depicts the variations of correlation coefficient with respect to the start time of each subregion.}
\label{fig:va_rw_cr}
\end{figure*}

For proper application of Wal\'{e}n test, the accurate estimation of background magnetic field is essential. However, it is not an observable quantity or it is not easily determined \citep{lichtenstein1979angle,riley1996properties,gosling2010torsional}. Generally, the average value of de Hoffmann-Teller (HT) frame or the mean value of the magnetic field is used for the complete duration of observations \citep{yang2013alfven,gosling2010torsional}.
However, HT frame can change quite fast in high-speed solar wind streams, and it is not always appropriate to take the average value of the magnetic field to be the background field \citep{li2016new}.
Gosling et al. (2009) ascribed that the solar wind fluctuations are pertinent to a slow varying base value rather than to an average value \citep{gosling2009one}. Further, they suggested that the conclusions derived from the analyses by assuming the fluctuations in all field components are relative to average values need to be reexamined. Thus, we thought that the observed degradation in Wal\'{e}n slopes and the correlation coefficients may cause by the varying background field since the distinct layers of scattered points are clearly evident (see Figure $4$ of Raghav et al. (2018)). Therefore, the complete green shaded region of Figure ~\ref{fig:IP} chopped in $15$ time-wise continuous sub-regions depending on the visible correlated layers of scattered points. The averaged values are estimated for each chopped layers and further re-analyzed by Wal\'{e}n test. Figure $2$ demonstrates relative fluctuations between $\Delta V_{Ax}$ and $\Delta V_x$, their correlation and linear relation for all 15 subregions (See attachment for variations in y- and z- components). The derived values of the amplitude ratio ($R_{V_A}$), Wal\'{e}n slope ($R_W$) and correlation coefficients of each component for various sub-regions with their start and stop times are shown in Table $1$. Their temporal variations are demonstrated in Figure ~\ref{fig:va_rw_cr}.

\begin{table*}
\begin{center}

\begin{tabular}{| c| c| c |c |c| c| c| c| c|}
\hline
\hline
Start time 	 & End	Time  & 	$X_{slope}$	 & 	$Y_{slope}$	 & 	$Z_{slope}$	 & 	mean $B_x$	 & 	$R_x$	 & 	$R_y$	 & 	$R_z$\\
\hline
\hline

18.8321	 & 	19.1261	 & 	2.4	 & 	0.64	 & 	0.6	 & 	-12.31	 & 	0.71	 & 	0.76	 & 	0.69\\
\hline
19.1261	 & 	19.4252	 & 	0.92	 & 	0.82	 & 	0.81	 & 	-10.98	 & 	0.77	 & 	0.94	 & 	0.92\\
\hline
19.4252	 & 	19.5887	 & 	0.46	 & 	0.63	 & 	0.99	 & 	-7.84	 & 	0.81	 & 	0.50	 & 	0.86\\
\hline
19.5887	 & 	19.736	 & 	0.064	 & 	0.66	 & 	0.57	 & 	-7.35	 & 	0.10	 & 	0.75	 & 	0.64\\
\hline
19.736	 & 	20.1156	 & 	-0.087	 & 	0.35	 & 	0.48	 & 	-5.27	 & 	-0.18	 & 	0.41	 & 	0.73\\
\hline
20.1156	 & 	20.3561	 & 	0.56	 & 	0.52	 & 	0.77	 & 	-0.53	 & 	0.79	 & 	0.58	 & 	0.90\\
\hline
20.3561	 & 	20.4355	 & 	0.4	 & 	0.22	 & 	0.31	 & 	-0.46	 & 	0.72	 & 	0.58	 & 	0.60\\
\hline
20.4355	 & 	20.6668	 & 	0.5	 & 	0.23	 & 	0.54	 & 	0.53	 & 	0.52	 & 	0.21	 & 	0.72\\
\hline
20.6668	 & 	20.9878	 & 	0.86	 & 	0.72	 & 	0.72	 & 	1.37	 & 	0.88	 & 	0.67	 & 	0.92\\
\hline
20.9878	 & 	21.2644	 & 	1.1	 & 	0.86	 & 	0.83	 & 	1.22	 & 	0.97	 & 	0.97	 & 	0.91\\
\hline
21.2644	 & 	21.3801	 & 	0.92	 & 	0.91	 & 	0.84	 & 	1.50	 & 	0.87	 & 	0.95	 & 	0.93\\
\hline
21.3801	 & 	21.498	 & 	0.91	 & 	0.98	 & 	0.97	 & 	0.75	 & 	0.98	 & 	0.92	 & 	0.95\\
\hline
21.498	 & 	21.5549	 & 	0.86	 & 	0.85	 & 	0.89	 & 	0.02	 & 	0.98	 & 	0.95	 & 	0.87\\
\hline
21.5549	 & 	21.7962	 & 	0.79	 & 	0.7	 & 	0.88	 & 	-2.35	 & 	0.93	 & 	0.95	 & 	0.98\\
\hline
21.7962	 & 	21.9143	 & 	0.83	 & 	0.84	 & 	0.95	 & 	1.93	 & 	0.98	 & 	0.93	 & 	0.93\\
\hline
\hline
\end{tabular}
\caption{The list of derived parameters}
\label{tab:1}
\end{center}
\end{table*}

\section{Results and Discussion }

An overall good correlation between magnetic field and proton velocity vectors is clearly evident in all aforementioned sub-divided regions. It indicates magnetic field and fluids are oscillating together, which confirms the presence of Alfv\'{e}n waves in the green shaded region of Figure ~\ref{fig:IP}. Particularly, 
in a magnetic cloud, correlation coefficient values vary from 0.81 to -0.18 for x-component, 0.94 to 0.21 for y-component and 0.92 to 0.64 for z-component. Similarly, Wal\'{e}n slopes varies from 2.4 to -0.087 for x-component, 0.82 to 0.22 for y-component and 0.99 to 0.31 for z-component. The lowest values of both parameters are seen after about $14:07~UT$ on 19 February onwards which further recovers. Moreover, the  explicit decrease in both values is again seen nearly about $08:32~UT$ on $20$ February. The \textit{in-situ} observations suggest the crossing of the magnetic cloud center by the Wind spacecraft around $\sim$ $12:00~ UT$ on $19$ February, whereas its end boundary is evident on $20$ February around $\sim$ $08:24~ UT$ (see Figure ~\ref{fig:IP}). The Wal\'{e}n slopes show sub-unity values for all components in a magnetic cloud, but after its boundary, both the values gradually recover to $\sim1$ and fluctuate close to 1. Our investigation suggests that the visual correlated layer-wise Wal\'{e}n test analysis improves the correlation coefficient and Wal\'{e}n slope values for the trailing solar wind region significantly.
 
The various physical mechanisms proposed to explain the observed sub-unity values for the Wal\'{e}n slope. For example, (i) the presence of minor ion species other than a proton in the solar wind \citep{puhl2000systematics}, (ii) the presence of compressive fluctuations \citep{bavassano1989evidence}, (iii) pickup ions through anisotropy \citep{goldstein1995alfven}, (iv) Alfv\'{e}n waves with mixed propagation directions \citep{belcher1971large}, (v) Electron flow velocities \citep{scudder1999generalized} and (vi) the acceleration of a rotational discontinuity \citep{sonnerup1987magnetopause,sonnerup1990magnetopause}  etc. Out of these, the mixture of inward and outward Alfv\'{e}n waves in the solar wind frame of reference \citep{d2015origin} is considered as the major cause of observed sub-unity values of Wal\'{e}n slope and could reduce the correlation between solar wind velocity and magnetic field \citep{belcher1971large,neugebauer2006comment,bruno2013solar,yang2016observational}. The outward Alfv\'{e}n waves are often observed in the solar wind, however inward Alfv\'{e}n waves are rarely seen \citep{belcher1969large,daily1973alfven,burlaga1976microscale,riley1996properties,denskat1982statistical,yang2016observational}. The inward Alfv\'{e}n waves are anticipated with increasing heliocentric distance and further linked with special events such as magnetic reconnection exhausts and/or back-streaming ions from reverse shocks \citep{belcher1971large,roberts1987origin,bavassano1989large,gosling2009one,gosling2011pulsed}. The wind velocity shears through plasma instabilities may cause superposition of inward and outward Alfv\'{e}n waves locally \citep{bavassano1989evidence}. The simultaneous presence of both the Alfv\'{e}n waves leads to non-linear interactions \citep{dobrowolny1980fully} which are not only crucial for the dynamical evolution of a Kolmogorov-like MHD spectrum \citep{bruno2013solar} but also for the depletion of the
normalized cross helicity of Alfv\'{e}nic fluctuations with
increasing heliocentric distance \citep{roberts1987origin,bavassano2000alfvenic,matthaeus2004transport}

For the studied event, Raghav \& Kule (2018) suggest that the Alfv\'{e}n wave is induced by the interaction of multiple CMEs which may change the force balance conditions of the fluxrope \citep{raghav2018first}. The 
Alfv\'{e}n wave could be generated by the steepening of a magnetosonic wave which forms the shock at the leading edge of the magnetic cloud\citep{tsurutani1988origin,tsurutani2011review}.  Besides this, the Alfv\'{e}n waves are commonly observed in interplanetary space \citep{hellinger2008oblique} but it would be difficult for them to get into the magnetic cloud. The high compression at the front edge of the magnetic cloud may be reflected in the first sub-divided region of x-component in which  unjustifiable Wal\'{e}n slope (very high $2.4$) is observed. The MHD static equilibrium suggests magnetic cloud/fluxrope as strongly coupled plasma under frozen in condition with concentric cylindrical surface layers around the central axis. The magnetic shear strength strongly couples these different concentric layers  \citep{russell1999magnetic}. Important to note that nearly after a half of magnetic cloud crossover and at the magnetic cloud end boundary, the Wal\'{e}n slope and correlation coefficient are minimum, further the $R_{V_A}$ values fluctuate close to $1$.

The distinct two types orientation in scattered points are observed in middle panel of the $2^{nd}$ row from the bottom of Figure $2$ which depicts the magnetic cloud end boundary region. In this region, the following solar wind is magnetically reconnecting with the end boundary of the magnetic cloud. Thus, magnetic reconnection process induces inward Alfv\'{e}n waves \citep{belcher1971large,roberts1987origin,bavassano1989large,gosling2009one,gosling2011pulsed} which may be responsible for the decrease in Wal\'{e}n slopes as well as correlation coefficients at the magnetic cloud boundary. The similar trend is also observed in a rightmost panel of $3^{rd}$ row from the bottom of Figure $2$. 
It implies the internal magnetic reconnection process within the magnetic cloud which leads to the simultaneous presence of inward and outward Alfv\'{e}n waves. The observations also imply that the plasma distribution in the x-direction (parallel) is highly affected as compare to the y- and z-direction (perpendicular). It would lead to induce pressure anisotropy within magnetic cloud which challenge the fundamental assumption \textit{i.e.} the thermal anisotropy is non-significant. This anisotropy in pressure can offsets
the magnetic tension and triggers the firehose instability on
ion gyroscales \citep{davidson1968macroscopic,yoon1993effect,hellinger2000new} which further leads to the interruption of Alfv\'{e}n wave. The simulation studies infer that the decay process of Alfv\'{e}n wave directly transfers large-scale mechanical energy
into thermal energy via viscous dissipation which heats the plasma \citep{squire2016stringent,squire2017amplitude}. Here, we demonstrate the observational signature in which the plasma temperature corresponding to the same region (where we have observed decrease in Wal\'{e}n slopes as well as correlation coefficients) gradually increases (see Figure ~\ref{fig:IP}) by approximately one order. 
Also,  the corresponding duration is nearly half a day which is comparable to Alfv\'{e}n wavelength (period). It means plasma gains the energy by some process which may be either caused by the magnetic reconnection and/or the dissipation of the Alfv\'{e}nic fluctuations in the corresponding regions.

We opine that this localized anisotropy affect magnetic shear strength and may trigger micro-scale instabilities \citep{de2016pressure} and sow the seeds for the internal magnetic reconnection between different concentric surface layers of magnetic cloud. The process can make the stable magnetic configuration unstable and may lead to tears down the magnetic cloud. 
Besides this, the $R_{V_A}$ values  within the magnetic cloud are observed well above $1$ in y- and z-component. It implies a dominant inward flow of Alfv\'{e}n waves which could be caused by their reflection  from the distinct magnetic boundaries of the cloud and energy remains confined within the cloud. The $R_{V_A}$ value  gradually decreasing below $1$ after the trailing edge of magnetic cloud crossover. It indicates dominant outward flow of Alfv\'{e}n waves along the following solar wind which disperses the energy in space.

The present study demonstrates the overall different characteristic of Alfv\'{e}n waves propagating in the magnetic cloud (dominant inward waves) and trailing solar wind (dominant outward waves). It has major implications in the dynamic evolution of the magnetic clouds (i.e. CMEs) which are the most stable structure in space and major drivers of heliospheric dynamics. The Sun is the main source of their ejection, further considering the number of CMEs ejected by the Sun, it is expected that the heliosphere should be full of  magnetic clouds.  However, no observational study support this expectation. It means there exists some process by which they tear down in the space and become part of the ambient solar wind. Moreover, how these magnetic structures disrupt in the space is still a mystery. The past studies suggest the CME-CME interactions formed complex ejecta in which the identity of individual CMEs (magnetic clouds) was lost \citep{burlaga2001fast,burlaga2002successive,farrugia2004evolutionary}, but how? is the question. Here, we proposed that the propagation of Alfv\'{e}n waves in magnetic cloud triggers instabilities in the stable magnetic configuration which leads to the disruption of magnetic clouds and further they become part of ambient solar wind. This might be a reason behind the rare observation of Alfv\'{e}n waves in magnetic cloud, despite the high expectation of such observations \citep{jacques1977momentum,gosling2010torsional}. Further studies are required to investigate the Alfv\'{e}n waves in interacting regions and their temporal evolution by multi-spacecraft measurements.

\section{Acknowledgements}
 We are thankful to WIND Spacecraft data providers (wind.nasa.gov) for  making interplanetary data available. We are also thankful to Department of Physics (Autonomous), University of  Mumbai, for providing us facilities for fulfillment of this work.  AR thanks to Solar-TErrestrial Physics (STEP) group, USTC, china \& SCOSTEP visiting Scholar program. AR also thankful to Siddharth Kasthurirangan for scientific discussions on disruption of Alfv\'{e}n wave.  

\bibliographystyle{mnras}

\begin{thebibliography}{}
	\makeatletter
	\relax
	\def\mn@urlcharsother{\let\do\@makeother \do\$\do\&\do\#\do\^\do\_\do\%\do\~}
	\def\mn@doi{\begingroup\mn@urlcharsother \@ifnextchar [ {\mn@doi@}
		{\mn@doi@[]}}
	\def\mn@doi@[#1]#2{\def\@tempa{#1}\ifx\@tempa\@empty \href
		{http://dx.doi.org/#2} {doi:#2}\else \href {http://dx.doi.org/#2} {#1}\fi
		\endgroup}
	\def\mn@eprint#1#2{\mn@eprint@#1:#2::\@nil}
	\def\mn@eprint@arXiv#1{\href {http://arxiv.org/abs/#1} {{\tt arXiv:#1}}}
	\def\mn@eprint@dblp#1{\href {http://dblp.uni-trier.de/rec/bibtex/#1.xml}
		{dblp:#1}}
	\def\mn@eprint@#1:#2:#3:#4\@nil{\def\@tempa {#1}\def\@tempb {#2}\def\@tempc
		{#3}\ifx \@tempc \@empty \let \@tempc \@tempb \let \@tempb \@tempa \fi \ifx
		\@tempb \@empty \def\@tempb {arXiv}\fi \@ifundefined
		{mn@eprint@\@tempb}{\@tempb:\@tempc}{\expandafter \expandafter \csname
			mn@eprint@\@tempb\endcsname \expandafter{\@tempc}}}
	
	\bibitem[\protect\citeauthoryear{Alfv{\'e}n}{Alfv{\'e}n}{1942}]{alfven1942existence}
	Alfv{\'e}n H.,  1942, Nature, 150, 405
	
	\bibitem[\protect\citeauthoryear{Bavassano \& Bruno}{Bavassano \&
		Bruno}{1989a}]{bavassano1989large}
	Bavassano B.,  Bruno R.,  1989a, Journal of Geophysical Research: Space
	Physics, 94, 168
	
	\bibitem[\protect\citeauthoryear{Bavassano \& Bruno}{Bavassano \&
		Bruno}{1989b}]{bavassano1989evidence}
	Bavassano B.,  Bruno R.,  1989b, Journal of Geophysical Research: Space
	Physics, 94, 11977
	
	\bibitem[\protect\citeauthoryear{Bavassano, Pietropaolo  \& Bruno}{Bavassano
		et~al.}{2000}]{bavassano2000alfvenic}
	Bavassano B.,  Pietropaolo E.,   Bruno R.,  2000, Journal of Geophysical
	Research: Space Physics, 105, 12697
	
	\bibitem[\protect\citeauthoryear{Belcher \& Davis}{Belcher \&
		Davis}{1971}]{belcher1971large}
	Belcher J.,  Davis L.,  1971, Journal of Geophysical Research, 76, 3534
	
	\bibitem[\protect\citeauthoryear{Belcher, Davis  \& Smith}{Belcher
		et~al.}{1969}]{belcher1969large}
	Belcher J.,  Davis L.,   Smith E.,  1969, Journal of Geophysical Research, 74,
	2302
	
	\bibitem[\protect\citeauthoryear{Boldyrev}{Boldyrev}{2006}]{PhysRevLett.96.115002}
	Boldyrev S.,  2006, \mn@doi [Phys. Rev. Lett.] {10.1103/PhysRevLett.96.115002},
	96, 115002
	
	\bibitem[\protect\citeauthoryear{Bruno \& Carbone}{Bruno \&
		Carbone}{2013}]{bruno2013solar}
	Bruno R.,  Carbone V.,  2013, Living Reviews in Solar Physics, 10, 2
	
	\bibitem[\protect\citeauthoryear{Burlaga}{Burlaga}{1971a}]{burlaga1971hydromagnetic}
	Burlaga L.,  1971a, Space Science Reviews, 12, 600
	
	\bibitem[\protect\citeauthoryear{Burlaga}{Burlaga}{1971b}]{burlaga1971nature}
	Burlaga L.~F.,  1971b, Journal of Geophysical Research, 76, 4360
	
	\bibitem[\protect\citeauthoryear{Burlaga \& Turner}{Burlaga \&
		Turner}{1976}]{burlaga1976microscale}
	Burlaga L.,  Turner J.,  1976, Journal of Geophysical Research, 81, 73
	
	\bibitem[\protect\citeauthoryear{Burlaga, Skoug, Smith, Webb, Zurbuchen  \&
		Reinard}{Burlaga et~al.}{2001}]{burlaga2001fast}
	Burlaga L.,  Skoug R.,  Smith C.,  Webb D.,  Zurbuchen T.,   Reinard A.,  2001,
	Journal of Geophysical Research: Space Physics, 106, 20957
	
	\bibitem[\protect\citeauthoryear{Burlaga, Plunkett  \& St~Cyr}{Burlaga
		et~al.}{2002}]{burlaga2002successive}
	Burlaga L.,  Plunkett S.,   St~Cyr O.,  2002, Journal of Geophysical Research:
	Space Physics, 107
	
	\bibitem[\protect\citeauthoryear{Chao, Hsieh, Yang  \& Lee}{Chao
		et~al.}{2014}]{chao2014walen}
	Chao J.,  Hsieh W.-C.,  Yang L.,   Lee L.,  2014, The Astrophysical Journal,
	786, 149
	
	\bibitem[\protect\citeauthoryear{Cramer}{Cramer}{2011}]{cramer2011physics}
	Cramer N.~F.,  2011, The physics of Alfv{\'e}n waves.
	John Wiley \& Sons
	
	\bibitem[\protect\citeauthoryear{D'Amicis \& Bruno}{D'Amicis \&
		Bruno}{2015}]{d2015origin}
	D'Amicis R.,  Bruno R.,  2015, The Astrophysical Journal, 805, 84
	
	\bibitem[\protect\citeauthoryear{Daily}{Daily}{1973}]{daily1973alfven}
	Daily W.~D.,  1973, Journal of Geophysical Research, 78, 2043
	
	\bibitem[\protect\citeauthoryear{Davidson \& V{\"o}lk}{Davidson \&
		V{\"o}lk}{1968}]{davidson1968macroscopic}
	Davidson R.~C.,  V{\"o}lk H.~J.,  1968, The Physics of Fluids, 11, 2259
	
	\bibitem[\protect\citeauthoryear{De~Camillis, Cerri, Califano  \&
		Pegoraro}{De~Camillis et~al.}{2016}]{de2016pressure}
	De~Camillis S.,  Cerri S.~S.,  Califano F.,   Pegoraro F.,  2016, Plasma
	Physics and Controlled Fusion, 58, 045007
	
	\bibitem[\protect\citeauthoryear{Denskat \& Neubauer}{Denskat \&
		Neubauer}{1982}]{denskat1982statistical}
	Denskat K.,  Neubauer F.,  1982, Journal of Geophysical Research: Space
	Physics, 87, 2215
	
	\bibitem[\protect\citeauthoryear{Dobrowolny, Mangeney  \& Veltri}{Dobrowolny
		et~al.}{1980}]{dobrowolny1980fully}
	Dobrowolny M.,  Mangeney A.,   Veltri P.,  1980, Physical Review Letters, 45,
	144
	
	\bibitem[\protect\citeauthoryear{Eastwood, Lucek, Mazelle, Meziane, Narita,
		Pickett  \& Treumann}{Eastwood et~al.}{2005}]{eastwood2005foreshock}
	Eastwood J.,  Lucek E.,  Mazelle C.,  Meziane K.,  Narita Y.,  Pickett J.,
	Treumann R.,  2005, Space Science Reviews, 118, 41
	
	\bibitem[\protect\citeauthoryear{Farrugia \& Berdichevsky}{Farrugia \&
		Berdichevsky}{2004}]{farrugia2004evolutionary}
	Farrugia C.,  Berdichevsky D.,  2004, in Annales Geophysicae. pp 3679--3698
	
	\bibitem[\protect\citeauthoryear{Foote \& Kulsrud}{Foote \&
		Kulsrud}{1979}]{foote1979hydromagnetic}
	Foote E.,  Kulsrud R.,  1979, The Astrophysical Journal, 233, 302
	
	\bibitem[\protect\citeauthoryear{Gizon, Cally  \& Leibacher}{Gizon
		et~al.}{2008}]{gizon2008helioseismology}
	Gizon L.,  Cally P.~S.,   Leibacher J.,  2008, Helioseismology,
	Asteroseismology, and MHD Connections.
	Springer Science \& Business Media
	
	\bibitem[\protect\citeauthoryear{Goldreich \& Sridhar}{Goldreich \&
		Sridhar}{1995}]{goldreich1995toward}
	Goldreich P.,  Sridhar S.,  1995, The Astrophysical Journal, 438, 763
	
	\bibitem[\protect\citeauthoryear{Goldstein, Neugebauer  \& Smith}{Goldstein
		et~al.}{1995}]{goldstein1995alfven}
	Goldstein B.,  Neugebauer M.,   Smith E.,  1995, Geophysical research letters,
	22, 3389
	
	\bibitem[\protect\citeauthoryear{Gosling, McComas, Roberts  \& Skoug}{Gosling
		et~al.}{2009}]{gosling2009one}
	Gosling J.,  McComas D.,  Roberts D.,   Skoug R.,  2009, The Astrophysical
	Journal Letters, 695, L213
	
	\bibitem[\protect\citeauthoryear{Gosling, Teh  \& Eriksson}{Gosling
		et~al.}{2010}]{gosling2010torsional}
	Gosling J.,  Teh W.-L.,   Eriksson S.,  2010, The Astrophysical Journal
	Letters, 719, L36
	
	\bibitem[\protect\citeauthoryear{Gosling, Tian  \& Phan}{Gosling
		et~al.}{2011}]{gosling2011pulsed}
	Gosling J.,  Tian H.,   Phan T.,  2011, The Astrophysical Journal Letters, 737,
	L35
	
	\bibitem[\protect\citeauthoryear{Heidbrink}{Heidbrink}{2008}]{heidbrink2008basic}
	Heidbrink W.,  2008, Physics of Plasmas, 15, 055501
	
	\bibitem[\protect\citeauthoryear{Hellinger \& Matsumoto}{Hellinger \&
		Matsumoto}{2000}]{hellinger2000new}
	Hellinger P.,  Matsumoto H.,  2000, Journal of Geophysical Research: Space
	Physics, 105, 10519
	
	\bibitem[\protect\citeauthoryear{Hellinger \&
		Tr{\'a}vn{\'\i}{\v{c}}ek}{Hellinger \&
		Tr{\'a}vn{\'\i}{\v{c}}ek}{2008}]{hellinger2008oblique}
	Hellinger P.,  Tr{\'a}vn{\'\i}{\v{c}}ek P.~M.,  2008, Journal of Geophysical
	Research: Space Physics, 113
	
	\bibitem[\protect\citeauthoryear{Hudson}{Hudson}{1971}]{hudson1971rotational}
	Hudson P.,  1971, Planetary and Space Science, 19, 1693
	
	\bibitem[\protect\citeauthoryear{Jacques}{Jacques}{1977}]{jacques1977momentum}
	Jacques S.,  1977, The Astrophysical Journal, 215, 942
	
	\bibitem[\protect\citeauthoryear{Li, Wang, Chao  \& Hsieh}{Li
		et~al.}{2016}]{li2016new}
	Li H.,  Wang C.,  Chao J.,   Hsieh W.,  2016, Journal of Geophysical Research:
	Space Physics, 121, 42
	
	\bibitem[\protect\citeauthoryear{Lichtenstein \& Sonett}{Lichtenstein \&
		Sonett}{1979}]{lichtenstein1979angle}
	Lichtenstein B.,  Sonett C.,  1979, Geophysical Research Letters, 6, 713
	
	\bibitem[\protect\citeauthoryear{Mari{\v{c}}i{\'c} et~al.,}{Mari{\v{c}}i{\'c}
		et~al.}{2014}]{marivcic2014kinematics}
	Mari{\v{c}}i{\'c} D.,  et~al., 2014, Solar physics, 289, 351
	
	\bibitem[\protect\citeauthoryear{Marsch}{Marsch}{2006}]{marsch2006kinetic}
	Marsch E.,  2006, Living Reviews in Solar Physics, 3, 1
	
	\bibitem[\protect\citeauthoryear{Matthaeus, Minnie, Breech, Parhi, Bieber  \&
		Oughton}{Matthaeus et~al.}{2004}]{matthaeus2004transport}
	Matthaeus W.~H.,  Minnie J.,  Breech B.,  Parhi S.,  Bieber J.,   Oughton S.,
	2004, Geophysical research letters, 31
	
	\bibitem[\protect\citeauthoryear{McIntosh, De~Pontieu, Carlsson, Hansteen,
		Boerner  \& Goossens}{McIntosh et~al.}{2011}]{mcintosh2011alfvenic}
	McIntosh S.~W.,  De~Pontieu B.,  Carlsson M.,  Hansteen V.,  Boerner P.,
	Goossens M.,  2011, Nature, 475, 477
	
	\bibitem[\protect\citeauthoryear{Mishra \& Srivastava}{Mishra \&
		Srivastava}{2014}]{mishra2014morphological}
	Mishra W.,  Srivastava N.,  2014, The Astrophysical Journal, 794, 64
	
	\bibitem[\protect\citeauthoryear{Neugebauer}{Neugebauer}{2006}]{neugebauer2006comment}
	Neugebauer M.,  2006, Journal of Geophysical Research: Space Physics, 111
	
	\bibitem[\protect\citeauthoryear{Ng \& Bhattacharjee}{Ng \&
		Bhattacharjee}{1996}]{ng1996interaction}
	Ng C.,  Bhattacharjee A.,  1996, The Astrophysical Journal, 465, 845
	
	\bibitem[\protect\citeauthoryear{Ofman}{Ofman}{2010}]{ofman2010wave}
	Ofman L.,  2010, Living Reviews in Solar Physics, 7, 4
	
	\bibitem[\protect\citeauthoryear{Puhl-Quinn \& Scudder}{Puhl-Quinn \&
		Scudder}{2000}]{puhl2000systematics}
	Puhl-Quinn P.,  Scudder J.,  2000, Journal of Geophysical Research: Space
	Physics, 105, 7617
	
	\bibitem[\protect\citeauthoryear{Quataert \& Gruzinov}{Quataert \&
		Gruzinov}{1999}]{quataert1999turbulence}
	Quataert E.,  Gruzinov A.,  1999, The Astrophysical Journal, 520, 248
	
	\bibitem[\protect\citeauthoryear{Raghav \& Kule}{Raghav \&
		Kule}{2018}]{raghav2018first}
	Raghav A.~N.,  Kule A.,  2018, \mn@doi [Monthly Notices of the Royal
	Astronomical Society: Letters] {10.1093/mnrasl/sly020}, 476, L6
	
	\bibitem[\protect\citeauthoryear{Raghav, Bhaskar, Lotekar, Vichare  \&
		Yadav}{Raghav et~al.}{2014}]{raghav2014quantitative}
	Raghav A.,  Bhaskar A.,  Lotekar A.,  Vichare G.,   Yadav V.,  2014, Journal of
	Cosmology and Astroparticle Physics, 2014, 074
	
	\bibitem[\protect\citeauthoryear{Raghav, Shaikh, Bhaskar, Datar  \&
		Vichare}{Raghav et~al.}{2017}]{raghav2017forbush}
	Raghav A.,  Shaikh Z.,  Bhaskar A.,  Datar G.,   Vichare G.,  2017, Solar
	Physics, 292, 99
	
	\bibitem[\protect\citeauthoryear{Raghav, Kule, Bhaskar, Mishra, Vichare  \&
		Surve}{Raghav et~al.}{2018}]{raghav2018torsional}
	Raghav A.~N.,  Kule A.,  Bhaskar A.,  Mishra W.,  Vichare G.,   Surve S.,
	2018, The Astrophysical Journal
	
	\bibitem[\protect\citeauthoryear{Riley, Sonett, Tsurutani, Balogh, Forsyth  \&
		Hoogeveen}{Riley et~al.}{1996}]{riley1996properties}
	Riley P.,  Sonett C.,  Tsurutani B.,  Balogh A.,  Forsyth R.,   Hoogeveen G.,
	1996, Journal of Geophysical Research: Space Physics, 101, 19987
	
	\bibitem[\protect\citeauthoryear{Roberts, Goldstein, Klein  \&
		Matthaeus}{Roberts et~al.}{1987}]{roberts1987origin}
	Roberts D.,  Goldstein M.,  Klein L.,   Matthaeus W.,  1987, Journal of
	Geophysical Research: Space Physics, 92, 12023
	
	\bibitem[\protect\citeauthoryear{Russell}{Russell}{1999}]{russell1999magnetic}
	Russell C.,  1999, Australian journal of physics, 52, 733
	
	\bibitem[\protect\citeauthoryear{Schlickeiser}{Schlickeiser}{2015}]{schlickeiser2015cosmic}
	Schlickeiser R.,  2015, Physics of Plasmas, 22, 091502
	
	\bibitem[\protect\citeauthoryear{Scudder, Puhl-Quinn, Mozer, Ogilvie  \&
		Russell}{Scudder et~al.}{1999}]{scudder1999generalized}
	Scudder J.,  Puhl-Quinn P.,  Mozer F.,  Ogilvie K.,   Russell C.,  1999,
	Journal of Geophysical Research: Space Physics, 104, 19817
	
	\bibitem[\protect\citeauthoryear{Sonnerup, Papamastorakis, Paschmann  \&
		L{\"u}hr}{Sonnerup et~al.}{1987}]{sonnerup1987magnetopause}
	Sonnerup B.,  Papamastorakis I.,  Paschmann G.,   L{\"u}hr H.,  1987, Journal
	of Geophysical Research: Space Physics, 92, 12137
	
	\bibitem[\protect\citeauthoryear{Sonnerup, Papamastorakis, Paschmann  \&
		L{\"u}hr}{Sonnerup et~al.}{1990}]{sonnerup1990magnetopause}
	Sonnerup B.,  Papamastorakis I.,  Paschmann G.,   L{\"u}hr H.,  1990, Journal
	of Geophysical Research: Space Physics, 95, 10541
	
	\bibitem[\protect\citeauthoryear{Squire, Quataert  \& Schekochihin}{Squire
		et~al.}{2016}]{squire2016stringent}
	Squire J.,  Quataert E.,   Schekochihin A.,  2016, The Astrophysical Journal
	Letters, 830, L25
	
	\bibitem[\protect\citeauthoryear{Squire, Schekochihin  \& Quataert}{Squire
		et~al.}{2017}]{squire2017amplitude}
	Squire J.,  Schekochihin A.,   Quataert E.,  2017, New Journal of Physics, 19,
	055005
	
	\bibitem[\protect\citeauthoryear{Temmer, Veronig, Peinhart  \&
		Vr{\v{s}}nak}{Temmer et~al.}{2014}]{temmer2014asymmetry}
	Temmer M.,  Veronig A.,  Peinhart V.,   Vr{\v{s}}nak B.,  2014, The
	Astrophysical Journal, 785, 85
	
	\bibitem[\protect\citeauthoryear{Tsurutani, Gonzalez, Tang, Akasofu  \&
		Smith}{Tsurutani et~al.}{1988}]{tsurutani1988origin}
	Tsurutani B.~T.,  Gonzalez W.~D.,  Tang F.,  Akasofu S.~I.,   Smith E.~J.,
	1988, Journal of Geophysical Research: Space Physics, 93, 8519
	
	\bibitem[\protect\citeauthoryear{Tsurutani, Lakhina, Verkhoglyadova, Gonzalez,
		Echer  \& Guarnieri}{Tsurutani et~al.}{2011}]{tsurutani2011review}
	Tsurutani B.,  Lakhina G.,  Verkhoglyadova O.~P.,  Gonzalez W.,  Echer E.,
	Guarnieri F.,  2011, Journal of Atmospheric and Solar-Terrestrial Physics,
	73, 5
	
	\bibitem[\protect\citeauthoryear{Wal{\'e}n}{Wal{\'e}n}{1944}]{walen1944theory}
	Wal{\'e}n C.,  1944, Arkiv for Astronomi, 30, 1
	
	\bibitem[\protect\citeauthoryear{Yang \& Chao}{Yang \&
		Chao}{2013}]{yang2013alfven}
	Yang L.,  Chao J.,  2013, Chin. J. Space Sci, 33, 353
	
	\bibitem[\protect\citeauthoryear{Yang, Lee, Chao, Hsieh, Luo, Li, Shi  \&
		Wu}{Yang et~al.}{2016}]{yang2016observational}
	Yang L.,  Lee L.,  Chao J.,  Hsieh W.,  Luo Q.,  Li J.,  Shi J.,   Wu D.,
	2016, The Astrophysical Journal, 817, 178
	
	\bibitem[\protect\citeauthoryear{Yoon, Wu  \& De~Assis}{Yoon
		et~al.}{1993}]{yoon1993effect}
	Yoon P.~H.,  Wu C.,   De~Assis A.,  1993, Physics of Fluids B: Plasma Physics,
	5, 1971
	
	\makeatother
\end{thebibliography}

\bsp	
\label{lastpage}
\end{document}